\documentclass[twocolumn,prb,prbbib,a4paper]{revtex4}
\usepackage{graphics}

\begin{document}

\title{\large
Tuning the onset voltage of resonant tunneling through InAs
quantum dots\\ by growth parameters}

\author{I.~Hapke-Wurst}
\author{U.~Zeitler\footnote{Present address:
Research Institute for Materials, High Field Magnet Laboratory, Toernooiveld 1, 6525 ED Nijmegen, The Netherlands}}
\author{U.~F.~Keyser}
\affiliation{
Institut f\"ur Festk\"orperphysik, Universit\"at Hannover,
Appelstr. 2, 30167 Hannover, Germany }
\author{K.~Pierz}
\author{Z.~Ma}
\affiliation{
Physikalisch-Technische Bundesanstalt Braunschweig, Bundesallee
100, 38116 Braunschweig, Germany}
\author{R.~J.~Haug}
\affiliation{
Institut f\"ur Festk\"orperphysik, Universit\"at Hannover,
Appelstr. 2, 30167 Hannover, Germany }

\date{\today}

\begin{abstract}
We investigated the size dependence of the ground state energy in
self-assembled InAs quantum dots embedded in resonant tunneling
diodes. Individual current steps observed in the current-voltage
characteristics are attributed  to resonant single-electron
tunneling via the ground state of individual InAs quantum dots.
The onset voltage of the first step observed is
shown to decrease systematically from 200 mV to 0
with increasing InAs coverage. We relate this
to a coverage-dependent size of InAs dots grown on AlAs. The
results are confirmed by atomic force micrographs and
photoluminescence experiments on reference samples.
\end{abstract}

\maketitle
\newlength{\plotwidth}          
\setlength{\plotwidth}{7.5cm}

Recently, self-assembled InAs quantum dots (QDs) grown in the
Stranski-Krastanov mode have attracted much interest as a novel
nanostructure with interesting electronic properties.
In particular, their well defined optical characteristics
as investigated in numerous photoluminescence experiments 
\cite{BimbergBuch} may be used for application in optical and 
opto-electronic devices such as QD laser and memories.~\cite{Application}

Until present only few experiments directly access the electronic transport
through InAs QDs by means of resonant tunneling through QDs embedded
single-barrier tunneling diodes.\cite{Nott1,Jap1,Jap2,Nott2,IHW1,IHW2}
Although typically 10$^5$  - 10$^6$ QDs are present in the
macroscopic devices used in such experiments, resonant single-electron
tunneling through individual QDs was observed.
This means that only a small proportion of the QDs in a
tunneling structure contribute to the tunneling current,
a much larger one is electrically inactive.
As a consequence, a correlation between the experimentally
observed single-QD features and the properties of the macroscopic
QD ensemble is far from being straightforward
in such type of experiments.

In this paper we will show that even in macroscopic diodes we can
nevertheless relate the onset voltage for resonant single-electron
tunneling through an individual QD to the ensemble properties of all QDs
which can be thoroughly controlled by growth parameters.
In particular, with increasing InAs coverage we find a systematically
decreasing onset voltage for the first
current step related to single-electron tunneling through the
energetically lowest lying electronically accessible InAs
QD. We relate this to an increasing quantum dot size with
increasing coverage leading to a lower ground state energy.
Our results are additionally supported by photoluminescence experiments and
atomic force micrographs on reference samples. These findings
may open up a new way to control the properties of
single-electron devices based on resonant tunneling via QDs.

The resonant-tunneling diode structures were grown by
molecular-beam epitaxy (MBE) on a 2 inch $n^+$-doped GaAs (100)
substrate ($n^+ = 2 \times 10^{24}$~m$^{-3}$).\cite{IHW1,IHW2}
A $n$-GaAs-back
contact of 1 $\mu$m thickness with graded Si-doping followed by an
undoped GaAs prelayer of 15 nm and a 5 nm AlAs barrier was
deposited at a substrate temperature $T_S = 600^o$C. During a
growth interruption of 60~s $T_S$ was lowered to 520$^o$C and the
substrate rotation was stopped. Subsequently, with a slow growth
rate of 0.05 ML/s, nominally 1.8 monolayers (ML) InAs, as measured in
the middle of the wafer, were deposited. The formation of self-assembled
QDs was observed by the change of the reflective high electron
energy diffraction (RHEED)
pattern after deposition of 1.6 ML of InAs.

Due to the geometric position of the In effusion cell with respect to
the non-rotating substrate an actual gradient ranging from 1.55 to
2.05 ML of InAs coverage is present across the wafer.
On the parts of the wafer where the InAs coverage $x$ exceeds 1.6 ML
we observe the formation of self-assembled QDs
in atomic force microscopy (AFM) images on comparable reference samples.
On the remaining parts of the wafer ($x < 1.6$~ML) only a
two-dimensional InAs wetting layer with no self-assembled QDs is
formed.

After the InAs growth and a subsequent growth interruption during 10 s
the wafer was rotated again and the QDs were covered with a 5 nm thick
AlAs barrier. Simultaneously $T_S$ was slowly ramped up to 600$^o$C.
Finally, another 15 nm of undoped GaAs and an upper contact
of 1 $\mu$m graded $n$-GaAs were grown.

Pieces with different InAs coverage were
cut from the wafer and tunneling diodes with standard AuGeNi contacts
were fabricated on a $40 \times 40~\mu$m$^2$ area by wet-chemically etching.
Each of these diodes contains approximately $5\times 10^9$ InAs QDs.
Reference wafers for AFM and
photoluminescence (PL) measurements were grown under similar
growth conditions. For the AFM experiments a wafer with uncovered QDs
grown on 20 nm of AlAs was used. An identical wafer, additionally capped
by 20 nm AlAs and 10 nm GaAs, was grown for the use in PL experiments.

\begin{figure}[t]
  \begin{center}
 \resizebox{7cm}{!}{\includegraphics{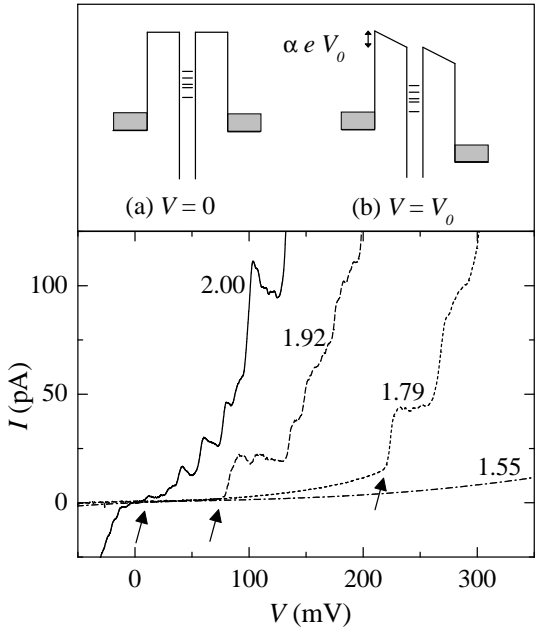}}
  \end{center}
  \caption{
  {\sl Top:} Schematic band structure of a tunneling diode at zero bias
             (a) and at finite bias where resonant tunneling through
             the first electronically accessible InAs QD sets in (b).
             The levels between the electrodes
             denote the energetic positions of the ground states of individual,
             electronically accessible InAs quantum dots.\protect{\newline}
  {\sl Bottom:} Current-voltage characteristics of samples with
                different InAs coverages. The numbers indicate the nominal
                number of monolayers of InAs deposited.
                The arrows mark the position of the first current steps occurring
                whenever resonant tunneling through the energetically lowest
                lying InAs QD sets in (sketch (b) on the top panel).}
  \label{IU}
\end{figure}

The layer structure of the resonant-tunneling diodes is designed
such that a 3D electron gas with a Fermi energy $E_F$ of
about 15 meV above the conduction band is formed in the GaAs
contact.\cite{IHW2} As sketched in the top panel of Fig.~\ref{IU}
the QD ground states are generally situated above $E_F$ at zero bias.
When increasing the bias voltage these states move towards lower energies
with respect to $E_F$, and, whenever $E_F$ comes into resonance with
such a QD level a step in the current-voltage ($I$-$V$)
characteristics appears, see Fig.~\ref{IU}(b).

In the bottom panel of Fig.~\ref{IU} we show such typical low-bias
$I$-$V$-curves for diodes with different InAs coverages
measured at 4.2~K.
In the structure with the lowest InAs coverage of 1.55 ML no QDs
have formed and,
consequently, a structureless, roughly exponential increase of the tunneling
current is observed. In contrast,
diodes with InAs coverages of more than 1.6 ML display step-like
increasing $I$-$V$-characteristics which we assign to the resonant
tunneling through individual InAs QDs. The first step in each $I$-$V$-curve
appears at an onset voltage $V_0$, marked by the arrows in Fig.~\ref{IU}. With
increasing InAs coverage $V_0$ systematically moves down towards
zero. This shows that the energy levels of the
QDs accessed by tunneling continuously shift to lower energies
when the InAs coverage is increased.\cite{IHW}

The onset voltage $V_0$ is in fact directly related to the
ground state energy $E_0$ of the first dot accessible to resonant
tunneling, $\alpha e V_0 = E_{0}$, where $E_0$ is measured with
respect to the Fermi energy in the emitter at zero bias. The
lever factor $\alpha $  accounts for the fact that only a part of
the total voltage applied drops between the emitter and the QD.
Since we observe an abrupt onset of current steps at a finite voltage
it is reasonable to assume that the first tunneling steps can be related
to dots with a relatively low ground state energy situated in the
low-energy tail of the QD energy distribution.
Subsequent steps then address dots with an energetically higher lying
ground state.

The total number of dots per diode that participate in resonant tunneling 
can be roughly estimated by referring the total current at higher voltages 
to the current contribution of one single dot. 
Supposing a current contribution of 20~pA for transport through one single dot
a current of typically 1~nA observed around 400~mV yields a total number of
about 50 dots electronically active at this bias voltage.
Each dot contibutes to the current in a bias range of 40~mV.\cite{Nauen}
Using the total energetic width of the dot distribution as deduced from
photoluminescence measurements (see below) we can extrapolate the 50 dots 
observed in a 40~mV range to an upper bound of totally less than 1000 dots 
participating in resonant tunneling over the entire bias voltage range.

\begin{figure}[t]
\begin{center}
 \resizebox{6cm}{!}{\includegraphics{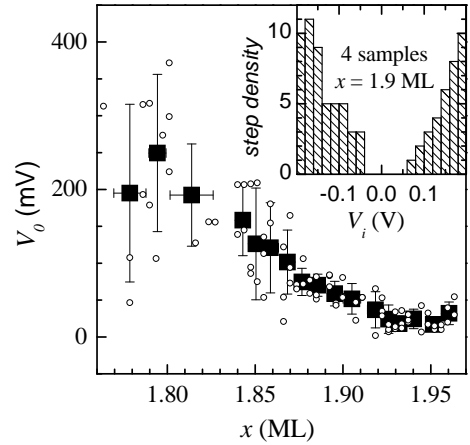}}
 \end{center}
  \caption{Onset voltage $V_0$ of the first InAs QD accessible
           to tunneling as a function of the InAs coverage.
           The small open dots correspond to individual samples.
           The solid rectangles represent an averaging over five
           samples with similar InAs coverage, the error bars show
           the statistical error of the average.
           \protect{\newline}
           The inset shows a histogram of the step density, 
           defined as the number of steps
           observed in a 20-mV interval, summed up over 4 samples 
           with $x = 1.9$ ML,
           as a function of the corresponding onset voltages.}
\label{Uon}
\end{figure}

The systematic decrease of $V_0$ with increasing InAs coverage
was verified by measuring totally 85 samples with InAs QDs.
All the measured onset voltages are compiled in Fig.~\ref{Uon}.
In particular for InAs coverages $x>1.8$~ML a strong correlation
between the onset voltage of the first current step and the InAs
coverage is found.
Such a correlation clearly shows that $V_0$ is strongly related
to the macroscopic ensemble properties of the InAs quantum dots,
and, in particular, to ground state energies $E_0$ decreasing
with increasing coverage.

\begin{figure}
\begin{center}
 \resizebox{5.2cm}{!}{\includegraphics{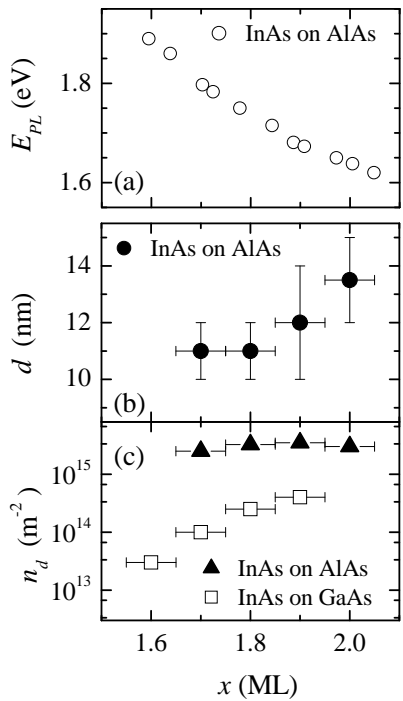}}
 \end{center}
   \caption
    {(a) Energy of the PL peak assigned to the optical transition inside
    the InAs QDs as a function of the InAs coverage.
    \protect{\newline}
    (b) Average dot size of InAs QDs grown on AlAs.
    \protect{\newline}
    (c)InAs-QD density as a function of the InAs coverage for dots grown on
          AlAs (closed triangles) and on GaAs (open squares).
          The data are deduced from AFM experiments on reference samples. }
\label{AFM-PL}
\end{figure}

We have verified this statement with
photoluminescence (PL) experiments on reference samples
performed at $T = 6$~K under cw excitation with an Ar$^+$ laser
(514 nm). As shown in Fig.~\ref{AFM-PL}(a) the position of the
PL peak corresponding to the average energy of the
ground-state transitions inside the InAs
QDs indeed moves systematically towards lower energies when
increasing the InAs coverage.

We have additionally investigated the energy distribution of the QDs
accessed by resonant tunneling by systematically measuring
4 samples with a similar InAs coverage $x \approx 1.9$~ML over a broad voltage
range. As can been seen in the inset of Fig.~\ref{Uon}
the step density, defined as the number of steps in a 20-mV interval
summed up over all four samples, increases with an increasing absolute value
of the bias voltage.
This clearly proves that the first steps indeed do access the
low energy tail of the QD distribution and subsequent steps
access energetically higher lying states with energies moving
towards the maximum in the energy distribution of the QDs.
Supposing a simple Gaussian distribution with a half-width
$\sigma \approx 70~$meV~as deduced from PL
and a lever factor $\alpha = 0.34$ \cite{IHW2} we extrapolate that only totally
40 dots per structure could be involved in resonant tunneling,\cite{rem}
about 8 of them are situated in the low energy
tail corresponding to onset voltages up to 200 mV.

The key point to understand the in Fig.~\ref{Uon} observed
systematic tendency in the coverage dependence of the onset
voltage for the first current step lies in the
fact that the size of the dots grown on relatively rough AlAs
is increasing when the InAs coverage is risen.\cite{InAs-AlAs}
In contrast, the size of InAs
QDs grown on a much smoother GaAs surface does merely depend on
the amount of InAs deposited.
Such a size variation of the InAs QDs as a function of the InAs coverage
can be verified independently by AFM, see Fig.~\ref{AFM-PL}(b).
Indeed, the lateral QD
dimensions increase from 10 to 14 nm in diameter when varying the
InAs coverage from 1.6 ML to 2.0 ML.
On the other hand, as shown in
Fig.~\ref{AFM-PL}(c), the QD density of InAs dots grown on AlAs,
$n_{QD} \approx 3 \times 10^{15}$m$^{-2}$ , is nearly
independent on the InAs coverage $x$,
whereas it strongly depends on $x$ for InAs QDs grown on GaAs.
This observation shows that the growth
of InAs QDs on AlAs (at the temperatures used in our experiments)
is kinetically limited by a slower In adatom diffusion compared to
the GaAs surface, which results into a fast nucleation of QDs
at positions predefined by the surface roughness.

In conclusion we have shown that the onset of resonant
single-electron tunneling through self-assembled InAs QDs embedded
in macroscopic tunneling diodes can be tuned between 0 and 0.3 V
by adjusting the amount of InAs deposited during MBE growth of the
corresponding device. This observation was explained by a
systematic increase of the InAs QDs with increasing InAs coverage
which can be related to the specific growth mechanism of QDs on
AlAs. The energetic distribution of the dots involved in resonant
tunneling was shown to correspond to that of the macroscopic ensemble.
However, only a small relative proportion of InAs QDs were shown to
participate in resonant tunneling, the rest are electronically inactive.


\begin{references}

\bibitem{BimbergBuch}
see e.g.: D.~Bimberg, M.~Grundmann, N.~N.~Ledentsov,
{\sl Quantum Dot Heterostructures} (Wiley, Chichester, 1999).

\bibitem{Application}
E.~Biolatti, I.~D'Amico, P.~Zanardi, and F.~Rossi,
Phys.~Rev.~B {\bf 65}, 075306 (2002) and references therein.


\bibitem{Nott1}
I.~E.~Itskevich, T.~Ihn, A.~Thornton, M.~Henini, T.~J.~Foster,
P.~Moriarty, A.~Nogaret, P.~H.~Beton, L.~Eaves, and P.~C.~Main,
Phys.~Rev.~{\bf B 54}, 16401 (1996).

\bibitem{Jap1}
T.~Suzuki, K.~Nomoto, K.~Taira, and I.~Hase,
Jpn.~J.~Appl.~Phys.~{\bf 36}, 1917 (1997).

\bibitem{Jap2}
M.~Narihiro, G.~Yusa, Y.~Nakamura, T.~Noda, and H.~Sakaki,
Appl.~Phys.~Lett.~{\bf 70}, 105 (1997).

\bibitem{Nott2}
A.~S.~G.~Thornton, T.~Ihn, P.~C.~Main, L.~Eaves, and M.~Henini,
Appl.~Phys.~Lett. {\bf 73}, 354 (1998).

\bibitem{IHW1}
I.~Hapke-Wurst, U.~Zeitler, H.~W.~Schumacher, R.~J.~Haug, K.~Pierz,
and F.~J.~Ahlers, Semicond.~Sci.~Technol.~{\bf 14}, L41 (1999).

\bibitem{IHW2}
I.~Hapke-Wurst, U.~Zeitler, H.~Frahm, A.~G.~M.~Jansen, R.~J.~Haug,
and K.~Pierz,
Phys.~Rev.~{\bf B 62}, 12621 (2000).

\bibitem{IHW}
I.~Hapke-Wurst, PhD thesis, Hannover (2002).

\bibitem{Nauen}
A. Nauen, I. Hapke-Wurst, F. Hohls, U. Zeitler, R.J. Haug, and K. Pierz,
Phys.~Rev.~{\bf B 66}, 161303(R), 2002.

\bibitem{InAs-AlAs}
P.~Ballet, J.~B.~Smathers, and G.~J.~Salamo,
Appl.~Phys.~Lett.~{\bf 75}, 337 (1999).

\bibitem{rem}
Even when supposing a considerably wider Gaussian energy distribution 
($\sigma=100$~meV) we find that than approximately totally 150 dots 
per structure would show resonant tunneling features, still only 
$3\times 10^{-5}$ of the total number of InAs QDs present in the structure.


\end{references}
\end{document}